\documentclass{jpp}
\usepackage{graphicx}
\usepackage{caption}
\usepackage{subcaption}
\usepackage{epstopdf, epsfig}
\usepackage{hyperref}

\usepackage{delarray}
\usepackage{here}
\usepackage{amsmath,amssymb}
\usepackage[latin1]{inputenc}
\usepackage{bm}
\usepackage{mathtools}
 \usepackage{xcolor}

\graphicspath{{Figures/}}

 \definecolor{red}{rgb}{1,0,0}
 \definecolor{gre}{rgb}{0,1,0}
 \definecolor{blu}{rgb}{0,0,1}

\newcommand{\be}{\begin{displaymath}}
\newcommand{\bn}{\begin{equation}}

\newcommand{\en}{\end{equation}}
\newcommand{\ee}{\end{displaymath}}
	\newcommand*\diff{\mathop{}\!\mathrm{d}}
\shorttitle{Stellarator boundary representation}
\shortauthor{Henneberg et al.}

\title{Representing the boundary of stellarator plasmas}

\author{S.A. Henneberg,
  \corresp{\email{sophia.henneberg@ipp.mpg.de}}
P. Helander  \and M. Drevlak }

\affiliation{Max Planck Institute for Plasma Physics, Greifswald, Germany}
\date{\today}

\begin{document}

\maketitle

\begin{abstract}

In stellarator optimization studies, the boundary of the plasma is usually described by Fourier series that are not unique: several sets of Fourier coefficients describe approximately the same boundary shape. A simple method for eliminating this arbitrariness is proposed and shown to work well in practice.

\end{abstract}

\section{Introduction}

In optimized stellarators, the magnetic field lines usually trace out simply nested flux surfaces. Large magnetic islands or regions with chaotic field lines are avoided, at least in the plasma core, in the interest of good confinement. Kruskal and Kulsrud have shown that magnetostatic equilibria with this property (insofar as they exist) are uniquely determined by the shape of the toroidal boundary and by the plasma current and pressure profiles \citep{Kruskal-1958,Helander-2014-a}. Instead of the current profile, that of the rotational transform can also be prescribed. 

This fundamental result provides the theoretical basis for fixed-boundary magnetohydrodynamic (MHD) equilibrium calculations, which are commonly used in stellarator optimization studies. The shape of the plasma boundary is prescribed, usually as a Fourier series in poloidal and toroidal angles \citep{Hirshman-1983,Nuehrenberg-1988}
	$$ R(\theta, \varphi) = \sum_{m=0}^M \sum_{n=-N}^N R_{m,n} \cos (m\theta - n \varphi), $$
	\bn Z(\theta, \varphi) = \sum_{m=0}^M \sum_{n=-N}^N Z_{m,n} \sin (m\theta - n \varphi), 
	\label{conventional representation}
	\en
and provides input to a fixed-boundary MHD equilibrium code. Here $(R,\varphi,Z)$ denote cylindrical coordinates and $\theta$ is a ``poloidal'' angle parameter, whose choice is the topic of this paper. For simplicity, we restrict our attention to fields with stellarator symmetry, 
	$$ R(\theta,\varphi) = R(-\theta,-\varphi), $$
	$$ Z(\theta,\varphi) =  - Z(-\theta,-\varphi). $$
Relinquishing this symmetry is not difficult, but the number of coefficients then needs to be doubled. 

In stellarator optimization, the Fourier coefficients $R_{m,n}$ and $Z_{m,n}$ are varied until an optimal magnetic equilibrium has been found, where the optimum is  defined by the minimum of some optimization target function. The optimization thus amounts to a search in a space of $2(M+1)(2N+1)$ dimensions.\footnote{Usually, the number is in fact slightly smaller, since negative values of $n$ are not included in the terms with $m=0$.} However, as has sometimes been remarked \citep{Hirshman-1998,Lee-1988}, this representation is not unique but contains ``tangential degrees of freedom'' in the limit $M \rightarrow \infty$, $N \rightarrow \infty$. If a large but finite number of terms are included in the Fourier series, several very different choices of the coefficients $\{R_{mn}, Z_{mn} \}$ correspond to approximately the same surface shape. Unless this problem is addressed, the search is therefore performed in a space of unnecessarily large dimensionality. Hirshman and co-workers devised a method called ``spectral condensation'' to deal with this problem, which is used internally in the VMEC and SPEC equilibrium codes to minimize the number of coefficients in the Fourier representation of all magnetic surfaces, including interior ones \citep{Hirshman-1983,Hirshman-1986-a,Hudson-2012-a}. However, spectral condensation is rarely used for the plasma boundary in optimization studies. The present article suggests another method of dealing with the problem of non-uniqueness of the boundary representation. This method is simpler but mathematically less sophisticated than spectral condensation. Unlike the latter, it does not correspond to a representation that is optimally economical, but it is simpler to implement numerically, requires less computation, and appears to work quite well in practice. 

The remainder of the present paper first describes the non-uniqueness of the representation (\ref{conventional representation}) and how it can be eliminated, followed by examples showing how this technique simplifies the problem of optimization by eliminating a plethora of spurious and approximate minima of the target function in configuration space. 

\section{Non-uniqueness of the usual representation}

In the representation (\ref{conventional representation}), the variable $\varphi$ denotes the toroidal geometric angle, but the choice of poloidal angle $\theta$ is arbitrary. Indeed, if we define
	$$ \tilde{R}(\theta, \varphi) = R(\theta + \epsilon(\theta,\varphi), \varphi), $$
	$$ \tilde{Z}(\theta, \varphi) = Z(\theta + \epsilon(\theta,\varphi), \varphi), $$	
where $\epsilon$ is any continuous, doubly $2\pi$-periodic function, then the surfaces 
	$$ S = \{(x,y,z) = (R(\theta,\varphi) \cos \varphi, R(\theta,\varphi)  \sin \varphi, Z(\theta,\varphi) ) : 0 \le \theta < 2 \pi, 0 \le \varphi < 2 \pi \} $$
and 
	$$ \tilde{S} = \{(x,y,z) = (\tilde R(\theta,\varphi) \cos \varphi, \tilde R(\theta,\varphi)  \sin \varphi, \tilde Z(\theta,\varphi) ) : 0 \le \theta < 2 \pi, 0 \le \varphi < 2 \pi \} $$
coincide. Moreover, if $|\p \epsilon(\theta,\varphi) / \p \theta| < 1$ for all $\theta$ and $\varphi$, then both surface parameterizations are bijective if one of them has this property. 

The fact that the addition of the function $\epsilon(\theta,\varphi)$ to the poloidal angle $\theta$ does not change $S$ indicates great freedom in the parameterization of the surface. If $M=N=\infty$ in the sum (\ref{conventional representation}), then infinitely many choices of coefficients $\{R_{mn}, Z_{mn}\}$ generate the same surface. Note that very different sets of coefficients can describe the same surface. If, on the other hand, $M$ and $N$ are finite in Eq.~(\ref{conventional representation}), so that the Fourier series of the functions $R(\theta,\varphi)$ and $Z(\theta,\varphi)$ terminate after a finite number of terms, then the corresponding series for $\tilde R(\theta,\varphi)$ and $\tilde Z(\theta,\varphi)$ will in general not terminate.\footnote{For simplicity, we take $\epsilon$ to satisfy $\epsilon(-\theta,-\varphi) = -\epsilon(\theta,\varphi)$ in order to preserve stellarator symmetry in the series (\ref{conventional representation}).}

This observation has implications for the nature of the representation (\ref{conventional representation}) when $M$ and $N$ are fixed, finite numbers:

\begin{enumerate}

\item[(i)] If $M$ and $N$ are not too large, so that only a few terms are kept in the series (\ref{conventional representation}), then the surface it represents will usually correspond to a unique set of coefficients $\{R_{m,n}, Z_{m,n}\}$ or a small number of such sets. Of course, with only a few harmonics, not every surface can be represented by Eq.~(\ref{conventional representation}). 

\item[(ii)] On the other hand, if many terms are included in the sum, $M \sim N \gg 1$, so that almost any stellarator-symmetric surface can be described the series Eq.~(\ref{conventional representation}), then many widely different choices of coefficients $\{R_{m,n}, Z_{m,n}\}$ can correspond to almost the same surface. The more harmonics that are allowed in the sum, the less unique the representation becomes.

These observations are confirmed by practical experience. At the beginning of a stellarator optimization run, it is usually futile to include many Fourier harmonics in the representation of the plasma boundary; the optimziation then ``gets stuck'' and does not proceed far from the initial state. Instead, it often proves useful to begin with only a few harmonics and gradually add more terms as the optimization proceeds, in order to allow for greater freedom in the shape of the plasma.

\section{Spectral condensation}
Spectral condensation exploits the non-uniqueness of the poloidal angle by minimizing the "spectral width", which measures the spectral extent of $R_{mn}$ and $Z_{mn}$, under the constraint of not changing the geometry of the surface $S$. The spectral width is defined by \cite{Hirshman-1985} and \cite{Hirshman-1998} as
$$ M\equiv \frac{\sum_{m,n} m^{(p+q)} (R_{m,n}^2 + Z_{m,n}^2)}{\sum_{m,n} m^p (R_{m,n}^2 + Z_{m,n}^2)}, $$
where $p \geq 0$ and $q >0$ are constants.\footnote{In the SPEC code \citep{Hudson-2012-a} it is defined without the normalization, i.e. 
$M\equiv \sum_{m,n} m^{(p+q)} (R_{m,n}^2 + Z_{m,n}^2). $} To first order in the perturbation $\epsilon$ introduced in the previous section, $\tilde R(\theta,\varphi) = R(\theta,\varphi) + \delta R(\theta,\varphi)$, $\tilde Z(\theta,\varphi) = Z(\theta,\varphi) + \delta Z(\theta,\varphi)$, and the Fourier coefficients of $\delta R$ and $\delta Z$ are given by
    $$ \delta R_{m,n} = \frac{1}{2\pi^2} \int \int \epsilon R_{\theta}
    \cos(m\theta - n \varphi) \diff \theta \diff \varphi, $$
    $$ \delta Z_{m,n} = \frac{1}{2\pi^2} \int \int \epsilon Z_{\theta}
    \sin(m\theta - n \varphi) \diff \theta \diff \varphi. $$
The first-order variation in the spectral width thus becomes
$$ \delta M = g^{-1} \int I(\theta,\varphi) \delta \epsilon(\theta,\varphi) \diff \theta \diff \varphi, $$
where
$$g = \pi^2 \sum_{m,n} m^p (R_{mn}^2 + Z_{mn}^2),$$
$$I(\theta,\varphi) =   X(\theta,\varphi) R_\theta + Y(\theta,\varphi) Z_\theta,$$
$$X (\theta,\varphi) =   \sum_{m,n} m^p(m^q-M) R_{m,n} \cos (m\theta-n\varphi),$$
$$Y (\theta,\varphi) =  \sum_{m,n} m^p(m^q-M) Z_{m,n} \sin (m\theta-n\varphi).$$
Note that the $m=0$ terms do not contribute to the Fourier sums, and that the spectral width assumes its minimum when $I(\theta,\varphi) =  0$. This constraint is imposed to the requisite accuracy by Fourier expanding $I(\theta,\varphi)$ and requiring a number $m^\ast$ of the Fourier coefficients $I_{mn}$ to vanish, thus effectively removing $m^\ast$ degrees of freedom from the representation \citep{Hirshman-1985}.

\section{An explicit boundary representation}
\label{Sec:explicitBR}
The method of spectral condensation is optimal in the sense that it minimizes the spectral width of the representation, but it adds conceptual and computational complexity. The number of coefficients in the representation (\ref{conventional representation}) remains high, although the constraints $I_{mn} = 0$ effectively restrict the search to a submanifold of lower dimensionality, and the system of equations corresponding to these constraints must in general be solved numerically. In this and the next section we explore a simpler and more explicit construction as a possible alternative. 

There are, of course, infinitely many ways of making the choice of poloidal angle unique, some of which have been proposed before, see e.g. \cite {Hirshman-1998} and \cite{Carlton-Jones-2020}.
A particularly simple choice could be to express the vertical coordinate as
    \bn Z(\theta, \varphi) = a(\varphi) + b(\varphi) \sin \theta. 
    \label{Z with two terms}
    \en
If the functions $a$ and $b$ are Fourier decomposed, 
    $$ a(\varphi) = \sum_{n=1}^N a_n \sin n\varphi, $$
    $$ b(\varphi) = \sum_{n=1}^N b_n \cos n\varphi, $$
one finds that this representation is of the same form as Eq.~(\ref{conventional representation}) but with only two poloidal harmonics,
	\bn Z(\theta, \varphi) = \sum_{m=0}^1 \sum_{n=-N}^N Z_{m,n} \sin (m\theta - n \varphi),
	\label{modified representation}
	\en
which are equal to 
    \bn Z_{0n} = - a_n, \qquad Z_{1n} = \frac{b_{|n|}}{2}. 
    \label{Zn}
    \en
	
\begin{figure}%[h!]
    \centering
    \includegraphics[width=0.5\textwidth]{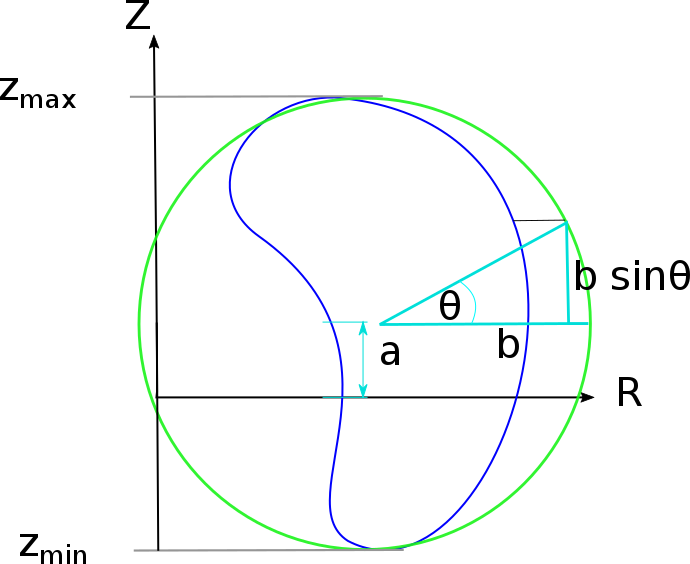}
    \caption{Circle (green) with radius $b=(z_{max}-z_{min})/2$, twice the total extent in the $Z$-direction of the plasma boundary (blue). The poloidal angle is chosen to be equal to the the polar angle of the horizontal projection of each boundary point on this circle.}
    \label{fig:rep1}
\end{figure}

In each poloidal cross section of the plasma surface, the poloidal angle $\theta$ thus defined is the polar angle of the horizontal projection on a circle with a diameter equal to the vertical extent of the surface, see Fig.~\ref{fig:rep1}. This choice of representation, which removes the superfluous degrees of freedom, can produce all surface shapes without multiple minima and maxima in the vertical coordinate $Z$ in each poloidal cross section. The vast majority of all stellarators considered to date possess this property.

However, Eq.~(\ref{modified representation}) suffers from another and more serious shortcoming: it cannot economically represent a classical stellarator. Such devices have an elliptical poloidal cross section that rotates co- or counter-clockwise with increasing toroidal angle $\varphi$. This can be seen by introducing a rotating coordinate system,
    $$ \rho = (R-R_0) \cos \alpha \varphi + Z \sin \alpha \varphi, $$
    $$ \zeta = - (R-R_0)  \sin \alpha \varphi + Z \cos \alpha \varphi, $$
where $\alpha$ is a constant determining the rate of rotation. For a classical stellarator, it is equal to half the number of toroidal periods of the device, $\alpha = N/2$, so that the cross section rotates by 180 degrees in one period.  In these coordinates, a surface with rotating elliptical boundary is represented by
    $$ \rho(\theta,\varphi) = A \cos \theta, $$
    $$ \zeta(\theta,\varphi) = B \sin \theta, $$
where $A$ and $B$ denote the semi-axes. In our original coordinates, we obtain
    $$ R(\theta,\varphi) = R_0 + \frac{A-B}{2} \cos (\theta - \alpha \varphi ) + \frac{A+B}{2} \cos (\theta + \alpha \varphi ), $$
    \bn Z(\theta,\varphi)= \frac{B-A}{2} \sin (\theta - \alpha \varphi ) + \frac{A+B}{2} \sin (\theta + \alpha \varphi ). 
    \label{Z}
    \en
Hence it is clear that $Z_{1,-1} \ne Z_{1,1}$ in contradiction to Eq.~(\ref{Zn}), which can only mean that the poloidal coordinate $\theta$ used in Eq.~(\ref{modified representation}) cannot coincide with the corresponding one in Eq.~(\ref{Z}). Although the former representation can describe {\em any} stellarator-symmetric surface, it needs {\em many} harmonics $R_{m,n}$ for a surface with rotating elliptical cross section. Close to the magnetic axis, most stellarators have this property, making this  shortcoming serious indeed. 

Fortunately, it is easily overcome by applying a representation similar to Eq.~(\ref{Z with two terms}) in the rotating coordinate system. This leads to the prescription
    $$ R(\theta,\varphi) = R_0(\varphi) + \rho(\theta, \varphi) \cos \alpha \varphi
    - \zeta(\theta, \varphi) \sin \alpha \varphi, $$
    $$ Z(\theta,\varphi) = Z_0(\varphi) + \rho(\theta, \varphi) \sin \alpha \varphi
    + \zeta(\theta, \varphi) \cos \alpha \varphi, $$
with
    $$ \rho(\theta, \varphi) = \sum_{m,n} \rho_{m,n} \cos(m\theta + n \varphi - \alpha \varphi), $$
    $$ \zeta(\theta, \varphi) = b(\varphi) \sin (\theta - \alpha \varphi) =
    \sum_{n=0}^N b_n \cos n\varphi \sin(\theta - \alpha \varphi), $$
which is our final recipe for unambiguously and  economically representing a stellarator-symmetric toroidal surface, see Fig.~\ref{fig:rep2}.
\begin{figure}%[h!]
    \centering
    \includegraphics[width=0.7\textwidth]{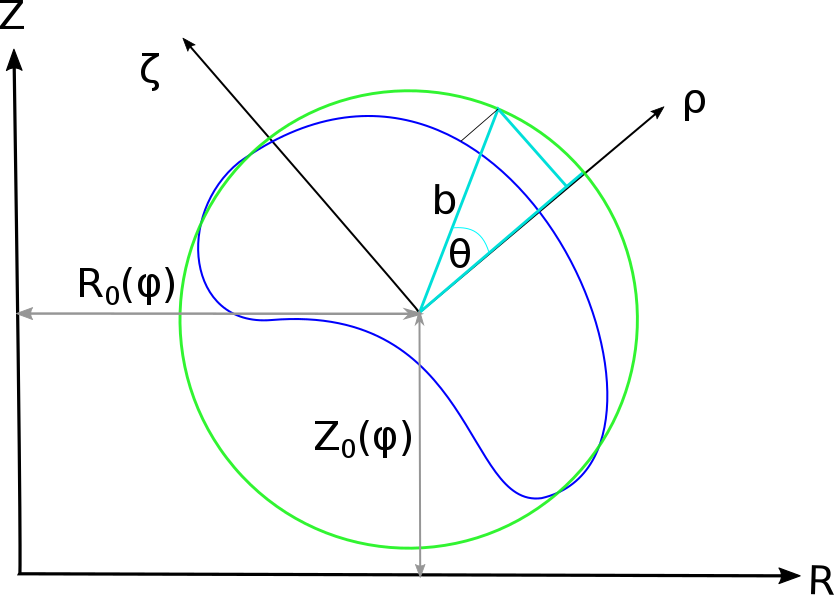}
    \caption{Circle (green) with radius $b=(\zeta_{max}-\zeta_{min})/2$ and the boundary (blue). The poloidal angle $\theta$ is the polar angle of the projection in the $\rho$-direction onto the circle.}
    \label{fig:rep2}
\end{figure}
In terms of our original representation (\ref{conventional representation}), the coefficients become for $m\neq 0$
    \bn R_{m,n} = \frac{1}{2} \left( \rho_{m,-n+2\alpha} + \rho_{m,-n} \right) + \frac{\delta_{m1}}{4}\left(b_n + b_{-n} - b_{n-2\alpha} - b_{-n+2\alpha} \right), \label{Rmn}
    \en
    \bn Z_{m,n} = \frac{1}{2} \left( \rho_{m,-n} - \rho_{m,-n+2\alpha} \right) + \frac{\delta_{m1}}{4}\left(b_n + b_{-n} + b_{n-2\alpha} + b_{-n+2\alpha} \right). \label{Zmn}
    \en
Specifically, if $\alpha=1/2$ we have
    \bn R_{m,n} = \frac{1}{2} \left( \rho_{m,-n+1} + \rho_{m,-n} \right) + \frac{\delta_{m1}}{4}\left(b_n + b_{-n} - b_{n-1} - b_{-n+1} \right), 
    \en
    \bn Z_{m,n} = \frac{1}{2} \left( \rho_{m,-n} - \rho_{m,-n+1} \right) + \frac{\delta_{m1}}{4}\left(b_n + b_{-n} + b_{n-1} + b_{-n+1} \right). 
    \en

In our experience, and as we shall see in the next section, this simple prescription works very well in practice.

\section{Numerical examples}

In this section, we explore a few examples of increasing complexity and realism, comparing our recipe with the conventional approach. 

\subsection{Simple axisymmetric (2D) case}

Our first aim is to gain insight into the optimization space using the original arbitrary-angle representation (\ref{conventional representation}). 
Since only the poloidal angle $\theta$ is arbitrary, this issue can be explored in a simpler, two-dimensional setting.
We choose to target an axisymmetric torus with a unit circle as the poloidal cross-section. A simple penalty function $Q$ that is minimized by this surface is
\begin{equation}
Q[R,Z] \coloneqq A^{-1}\oint  \left((R(\theta)-R_0)^2 +Z^2(\theta) - 1\right)^2 \diff s ,   \label{Q} 
\end{equation} 
where
$$\diff s = R\sqrt{(R_\theta^2 + Z_\theta^2)}\diff \theta \diff \varphi, $$
Here subscripts indicate partial derivatives, $R_\theta = \partial R/\partial \theta$,  
the surface area is $A\coloneqq \oint \diff s$, and the major radius is arbitrarily chosen to be $R_0=1.5$. We restrict $R$ and $Z$ to be axisymmetric: $R_{mn}=Z_{mn}=0$ for all $n\neq0$ and write
$$ R=1.5+\sum_{m=1} R_m  \cos(m \theta), $$
and
$$Z =\sum_{m=1} Z_m \sin(m \theta). $$
If $Q$ is not normalized to the area $A$, an artificial minimum exists when the area becomes small. With the normalization, the penalty function $Q$ approaches unity in the limit of vanishing $R-R_0$ and $Z$ (and thus vanishing surface area). The penalty function attains its sole minimum ($Q=0$) if $(R-R_0)^2 + Z^2 = 1$. 
This equation is satisfied by many different choices of $R_{m}$ and $Z_{m}$.\footnote{Although this is the only global minimum, $Q$ becomes arbitrarily small for bounded surfaces having very large area, e.g., for highly ``wrinkled'' surfaces.}

%Results
If all but the $m=1$ Fourier coefficients vanish, i.e. if $Q=Q(R_1,Z_1,R_i=0,Z_i=0)$ for $i>0$, the penalty function attains the global minimum for $R_1=Z_1=1$, see Fig.~\ref{fig:Q1}, and this is in general the case when only one pair of coefficients $(R_m,Z_m)$ is allowed to be non-zero, see Fig.~\ref{fig:Q2}.  
\begin{figure}%[h]
    \centering
    \includegraphics[width=.5\textwidth]{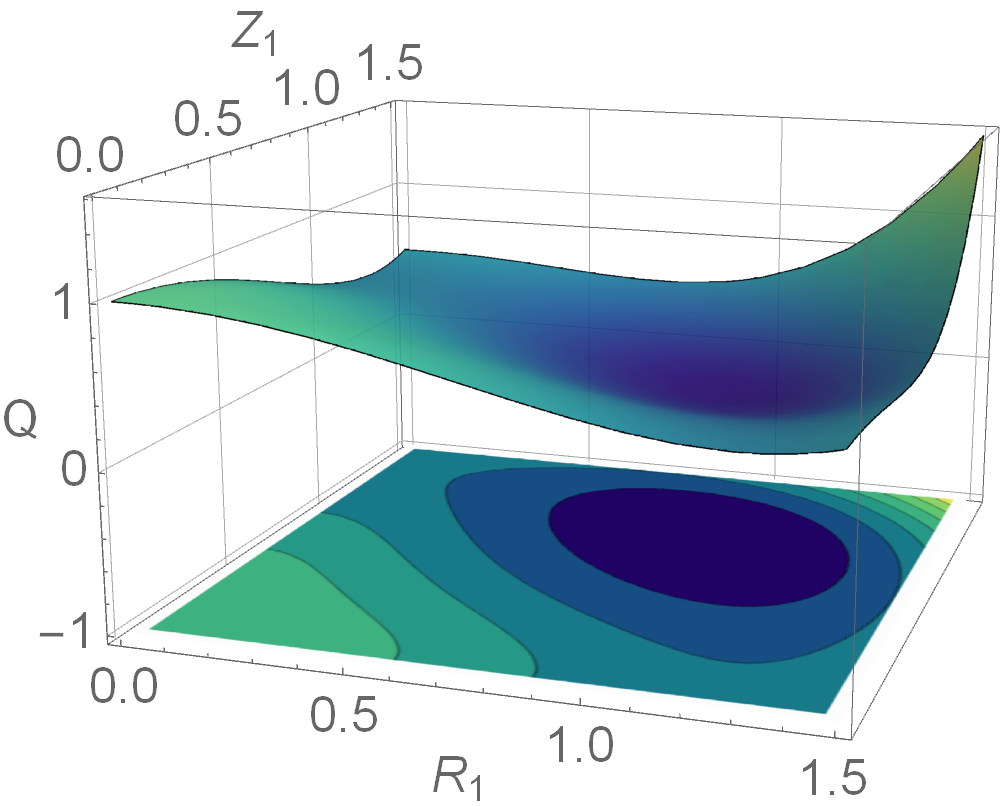}
    \caption{The cost function $Q(R_1,Z_1)$ with respect to $R_1$ and $Z_1$ with all other Fourier harmonics equal to zero $R_i=Z_i=0$, $i>1$.}
    \label{fig:Q1}
\end{figure}
\begin{figure}%[h]
    \centering
    \includegraphics[width=.5\textwidth]{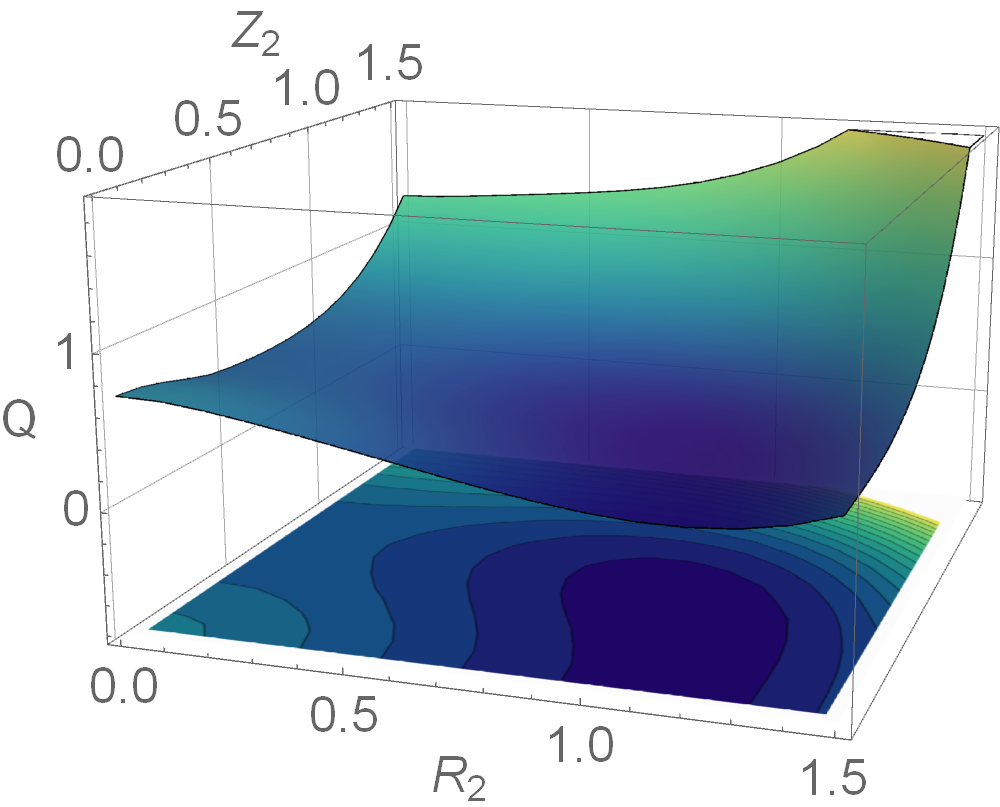}
    \caption{The cost function $Q(R_1=0.0,R_2,Z_1=0.7,Z_2)$ with respect to $R_2$ and $Z_2$.}
    \label{fig:Q2}
\end{figure}

More interesting and complex behaviour is observed if Fourier harmonics with several values of $m$ are admitted, but one then faces the problem of graphically displaying the function $Q$ of more than two variables. For instance, it is not easy to visualize how the cost function $Q(R_1,R_2,Z_1,Z_2)$ depends on all four arguments. However, some insight can be gained by plotting the mininum of $Q$ with respect to two of the arguments as a function of the two other ones, e.g. by considering the function
    $$ \tilde Q(R_1, Z_1) = \min_{R_2, Z_2} Q(R_1,R_2,Z_1,Z_2). $$

Considering four Fourier harmonics in this way reveals the existence of three local minima, see Fig.~\ref{fig:minQ1}. 
\begin{figure}%[h!]
    \centering
    \includegraphics[width=.7\textwidth]{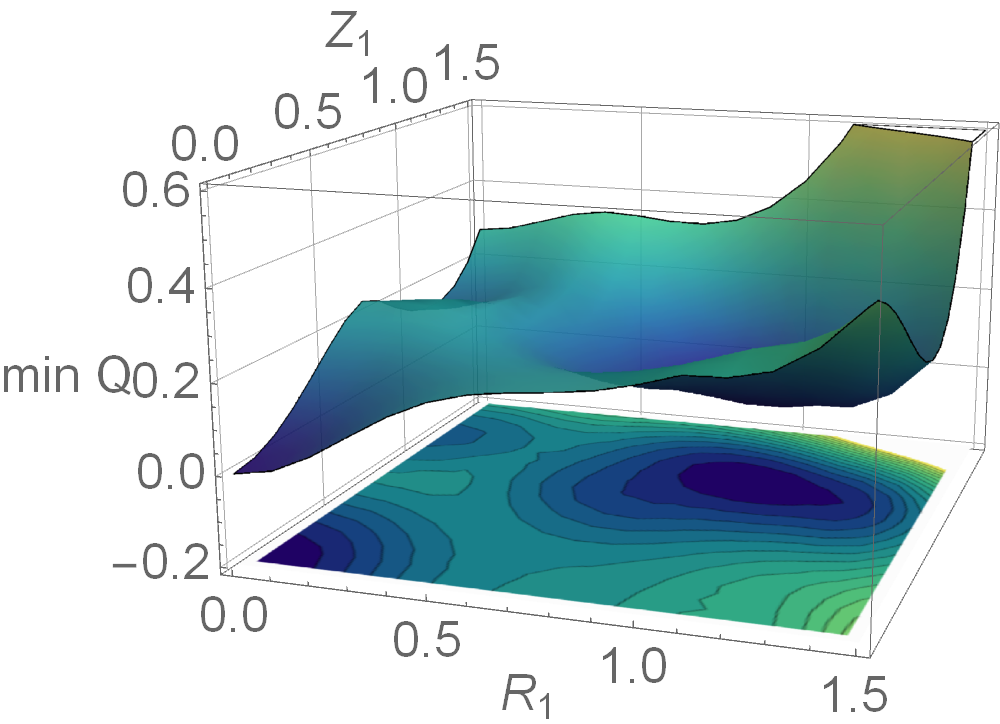}
    \caption{The minimized cost function $\min_{R_2,Z_2}Q(R_1,R_2,Z_1,Z_2)$ with respect to $R_1$ and $Z_1$.}
    \label{fig:minQ1}
\end{figure}
Two of these correspond to the global minimum, $R_1=Z_1=1, R_2= Z_2=0$ and $R_1= Z_1=0, R_2=Z_2=1$, and both correspond to the target surface, an axisymmetric torus with a unit circle cross section, parameterized in two different ways. The third minimum is located at $R_1=0.0, R_2\approx 1.05, Z_1=1.15$, and $Z_2=0.0$ with $Q\approx0.133$. It corresponds to a surface with zero volume but finite area, see Fig.~\ref{fig:minimaQ1}. 
\begin{figure}
\centering
\begin{subfigure}{.5\textwidth}
  \centering
  \includegraphics[width=.8\linewidth]{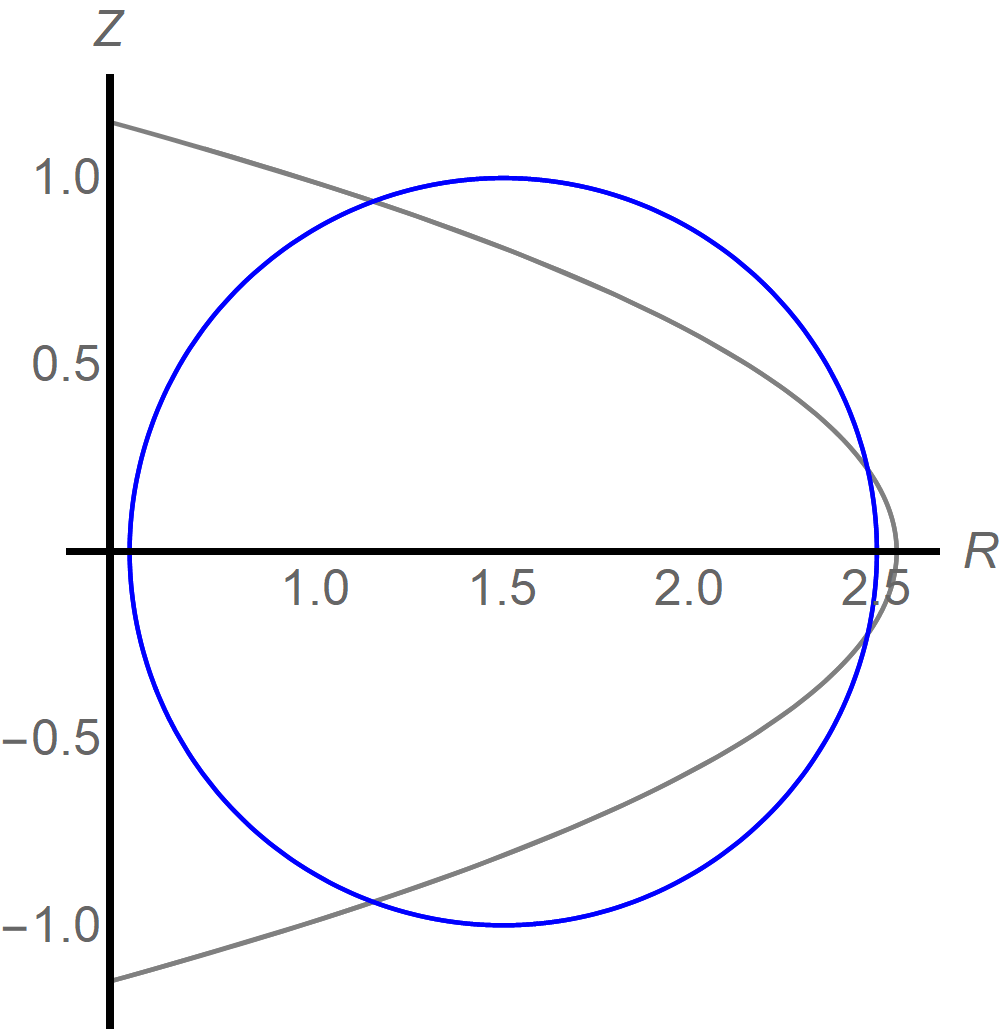}
  \caption{The poloidal cross section.}
  \label{fig:cross_section}
\end{subfigure}%
\begin{subfigure}{.5\textwidth}
  \centering
  \includegraphics[width=1.\linewidth]{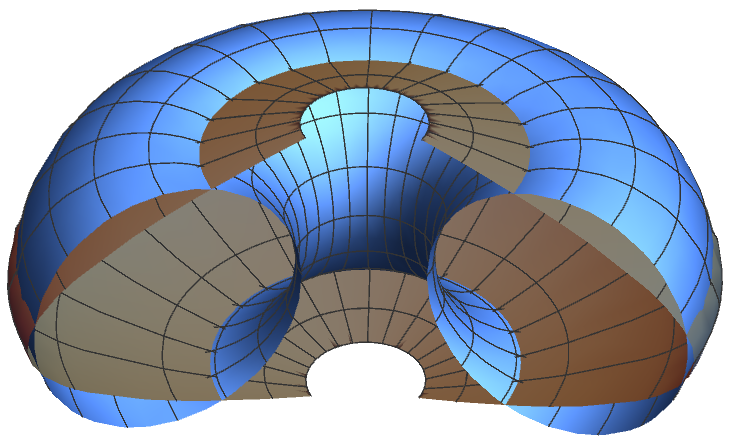}
  \caption{3D view.}
  \label{fig:3Dminima}
\end{subfigure}
\caption{The three minima of $Q(R_1,R_2,Z_1,Z_2)$. In blue (global minima): axisymmetric torus with unit circle cross section described by $R_1= Z_1=1, R_2= Z_2=0$ and $R_1=Z_1=0, R_2=Z_2=1$ ($R_i=0$ for all $i>2$). In gray (local minimum): $R_1=0.0, R_2\approx 1.05, Z_1=1.15$, and $Z_2=0.0$.}
\label{fig:minimaQ1}
\end{figure}

Increasing the number of poloidal harmonics $m$ to three makes it more difficult to locate the local minima of the cost function. Without a global optimizer, one encounters many local minima depending on the initial values chosen for the remaining Fourier coefficients. Using differential evolution, a global optimization routine, to obtain $\min_{R_2,Z_2,R_3,Z_3}Q(R_1,R_2,R_3,Z_1,Z_2,Z_3)$ one finds a landscape broadly similar to the one found for the four-Fourier-coefficient case, but with many additional small local maxima and minima, see Fig.~\ref{fig:minQ2}. This type of scan has to be considered with caution. Most global optimization routines do not, in practice, guarantee a global minimum but sometimes end up in local ones. In local optimization routines, this problem is of course still more acute, since the outcome generally depends on the initialization. 
\begin{figure}%[h!]
    \centering
    \includegraphics[width=.7\textwidth]{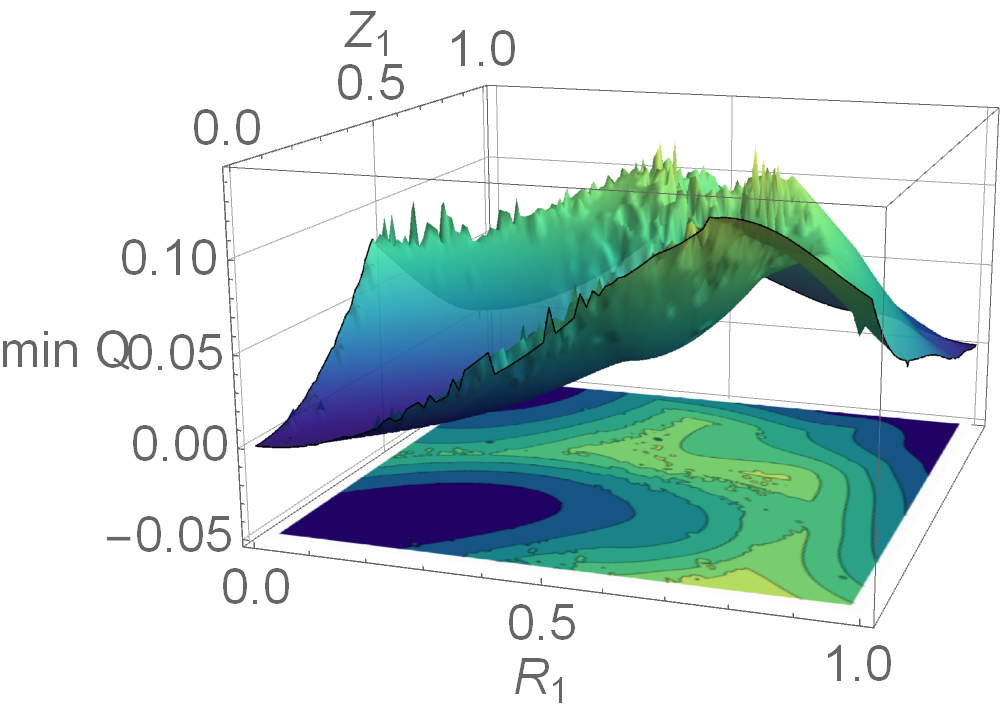}
    \caption{The minimized cost function $\min_{R_2,Z_2,R_3,Z_3}Q(R_1,R_2,R_3,Z_1,Z_2,Z_3)$ with respect to $R_1$ and $Z_1$. Differential Evolution, a global optimization routine, was used to find the minima.}
    \label{fig:minQ2}
\end{figure}

To visualize the difficulty of finding global minima, it is useful to fix two coefficients, in the following $R_1(=0.18)$ and $Z_1(=0.4)$, and study how the landscape depends on the remaining ones. We note that the function $\min_{R_3,Z_3} Q(R_1=0.18,R_2,R_3,Z_1=0.4,Z_2,Z_3)$ possesses five local minima with $R,2$ and $Z_2$ in the range  $0.0-1.0$, see Fig.~\ref{fig:minQ2_2} and line discontinuity, as can be seen in Fig.~\ref{fig:minQ2_2}. To understand the discontinuity $\min_{R_3,Z_3}Q(R_1=0.18,R_2,R_3,Z_1=0.4,Z_2,Z_3)$, we plot $Q(R_1=0.18,R_2=x,R_3,Z_1=0.4,Z_2=y,Z_3)$ the function $R_3$ and $Z_3$ for selected values for $x$ and $y$, Fig.~\ref{fig:Qr3z3}. The number of local minima varies with  $x$ and $y$. For $(x,y)=(0.24,0.39)$  there are four local minima in the figure, for $(x,y)=(0.5,0.5)$ there are two of them, and for $(x,y)=(1,1)$ there is only one minimum. It is thus clear that a local optimizer that seeks local minima of the function $Q(R_1=0.18,R_2,R_3,Z_1=0.4,Z_2,Z_3)$ will find different ones depending on the starting point for $R_3$ and $Z_3$. Abrupt changes (discontinuity) in $\min_{R_3,Z_3}Q(R_1=0.18,R_2,R_3,Z_1=0.4,Z_2,Z_3)$ appear when a local minimum  disappears and the optimizer finds a different one.    
\begin{figure}%[h!]
    \centering
    \includegraphics[width=.7\textwidth]{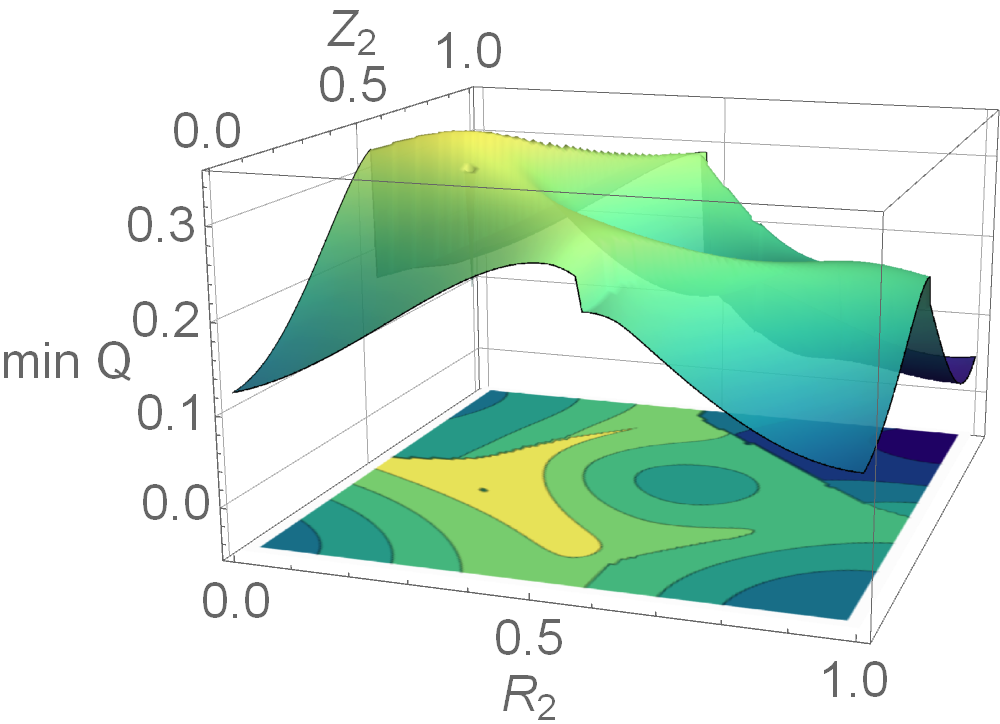}
    \caption{The locally minimized cost function $\min_{R_3,Z_3}Q(R_1=0.18,R_2,R_3,Z_1=0.4,Z_2,Z_3)$ as a function of $R_2$ and $Z_2$.}
    \label{fig:minQ2_2}
\end{figure}
\begin{figure}
\centering
\begin{subfigure}{.3\textwidth}
  \centering
  \includegraphics[width=1.\textwidth]{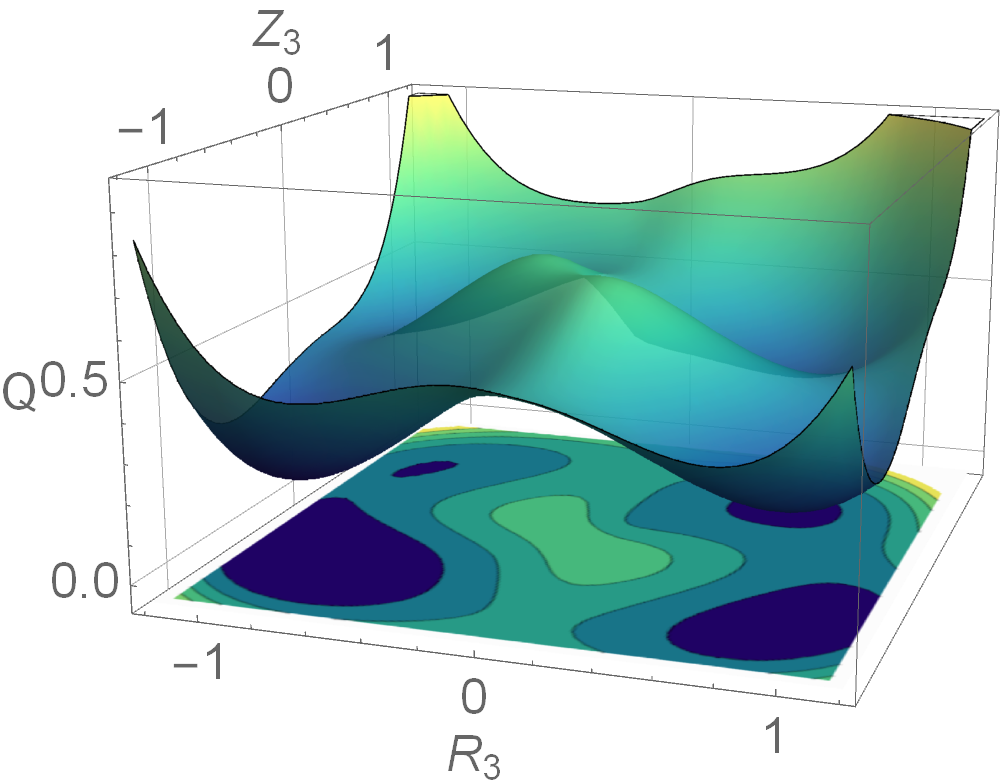}
  \caption{$x=0.24$ and $y=0.39$.}
  \label{fig:Qr3z3_1}
\end{subfigure}%
\begin{subfigure}{.3\textwidth}
  \centering
  \includegraphics[width=1.\textwidth]{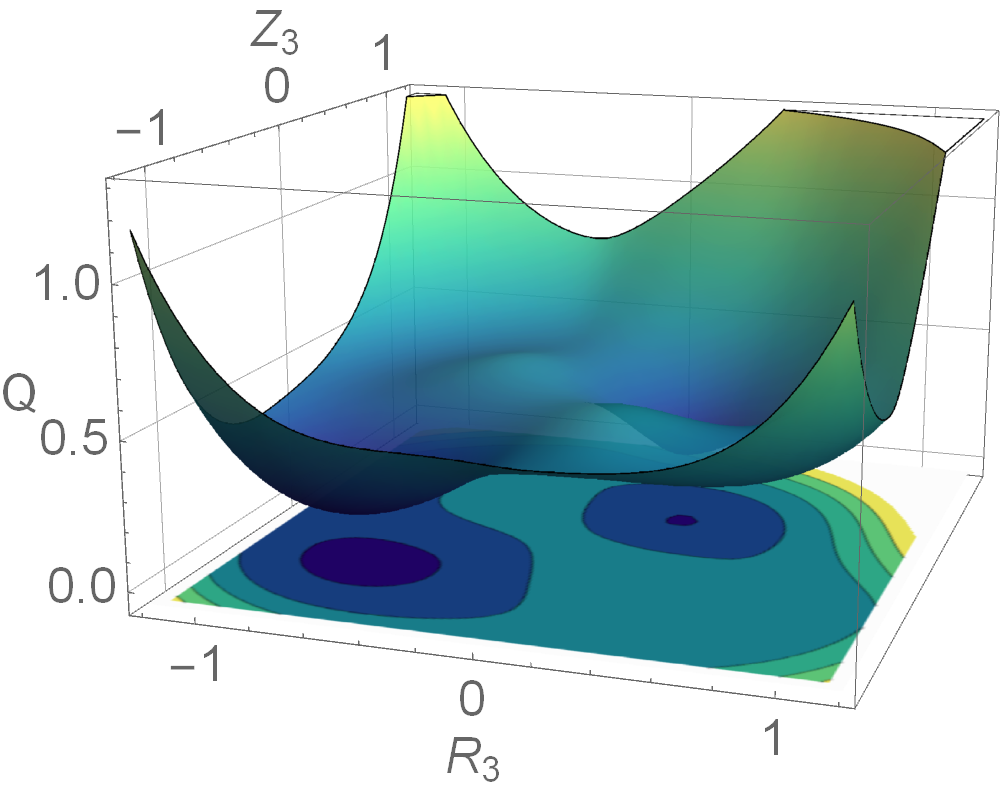}
  \caption{$x=0.5$ and $y=0.5$.}
  \label{fig:Qr3z3_2}
\end{subfigure}
\begin{subfigure}{.3\textwidth}
  \centering
  \includegraphics[width=1.\textwidth]{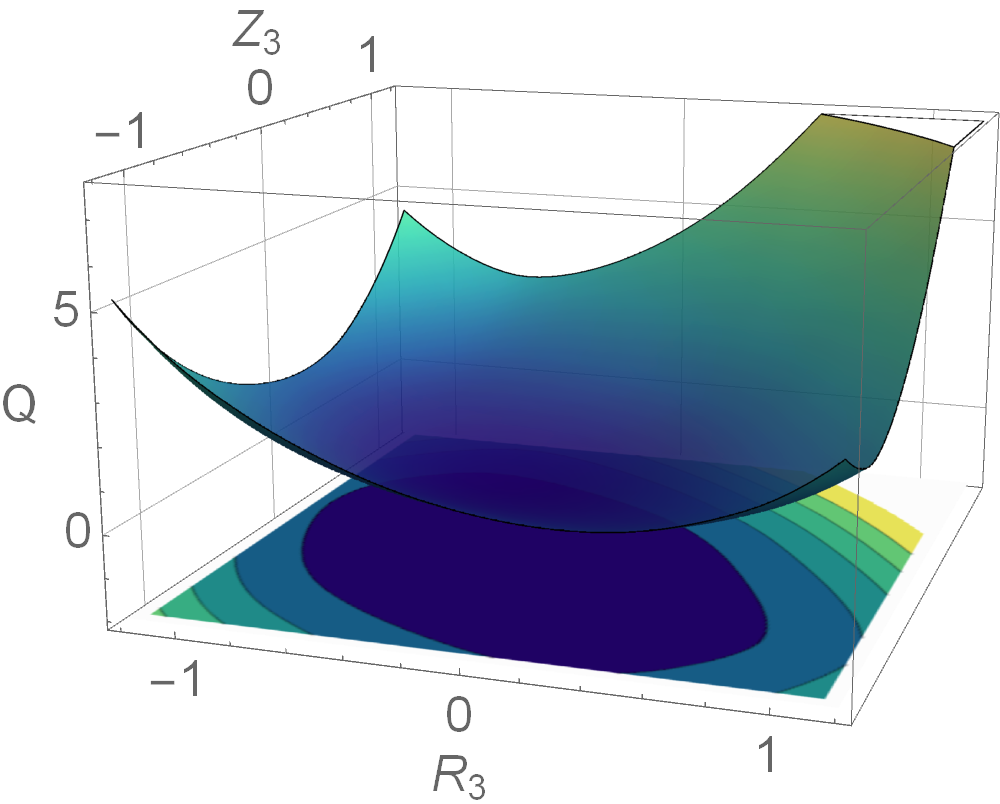}
  \caption{$x=1.0$ and $y=1.0$.}
  \label{fig:Qr3z3_3}
\end{subfigure}
\caption{$Q(R_1=0.18,R_2=x,R_3,Z_1=0.4,Z_2=y,Z_3)$ with respect to $R_3$ and $Z_3$.}
\label{fig:Qr3z3}
\end{figure}

The representation proposed in Sect.~\ref{Sec:explicitBR} leads to much more benign results when applied to the model problem (\ref{Q}). Restricting the optimization space to axisymmetric designs leads to
$$ R=1.5+\sum_{m=1} R_m  \cos(m \theta), $$
$$Z =Z_1 \sin(\theta). $$
This time, we find that the landscape of the minimum penalty function $\min Q$ does not change much when the number of Fourier harmonics of $R$ is increased. In the case of four Fourier harmonics, $Q(R_1,R_2,R_3,Z_1)$, there is one global minimum at $R_1=1,R_2=0,R_3=0,Z_1=1$ and a second shallow local minimum near $R_1=0$ and $Z_1=1.0$, see Fig.~\ref{fig:minQ_new}. This local minimum disappears when more Fourier harmonics are added.
\begin{figure}%[h!]
    \centering
    \includegraphics[width=.7\textwidth]{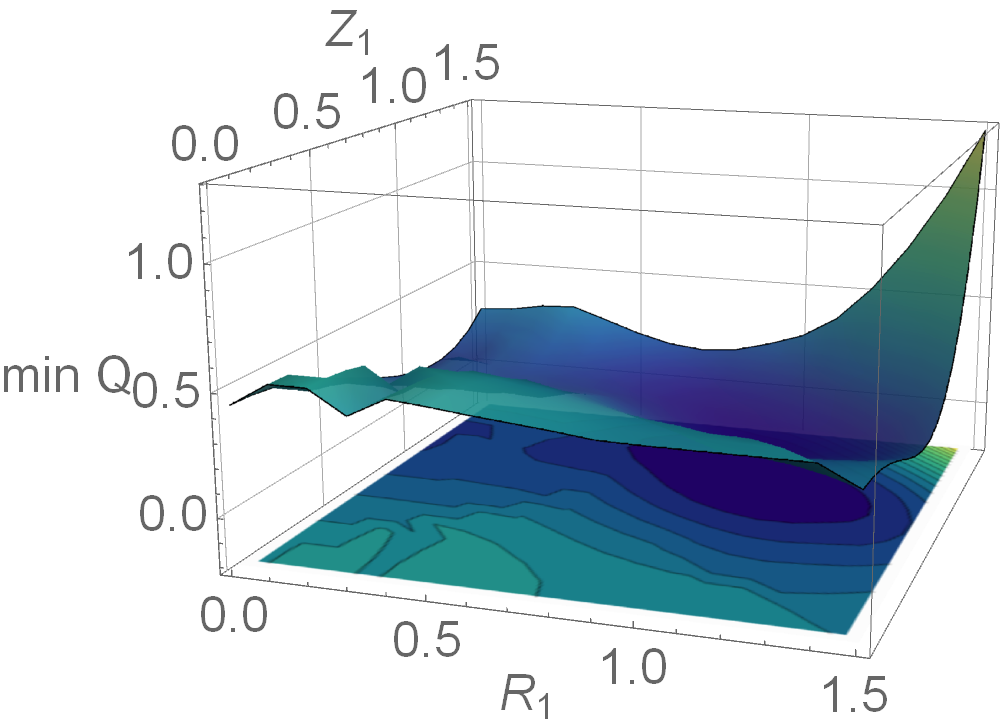}
    \caption{The minimized cost function $\min_{R_2,R_3}Q(R_1,R_2,R_3,Z_1)$ with respect to $R_1$ and $Z_1$.}
    \label{fig:minQ_new}
\end{figure}
Importantly, the outcome is similar whether a local and global optimization algorithm is employed, see Fig.~\ref{fig:New2_local} and Fig.~\ref{fig:New2_global}, making it much easier to find the minima numerically. 
\begin{figure}
\centering
\begin{subfigure}{.5\textwidth}
  \centering
  \includegraphics[width=1.\linewidth]{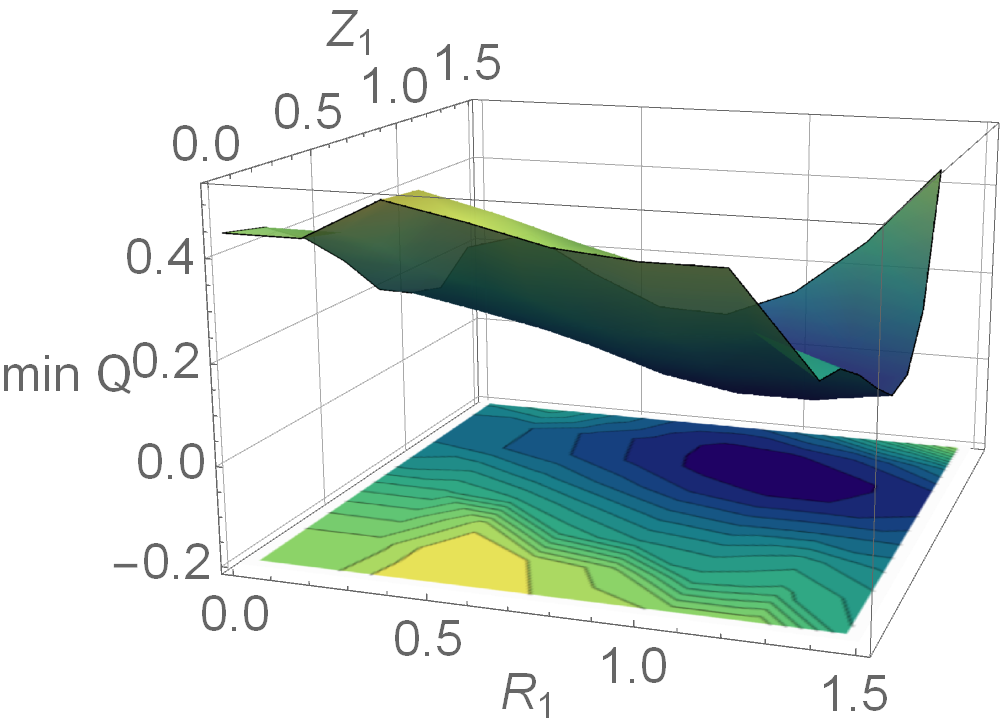}
  \caption{ using a non-global optimization algorithm.}
  \label{fig:New2_local}
\end{subfigure}%
\begin{subfigure}{.5\textwidth}
  \centering
  \includegraphics[width=1.\linewidth]{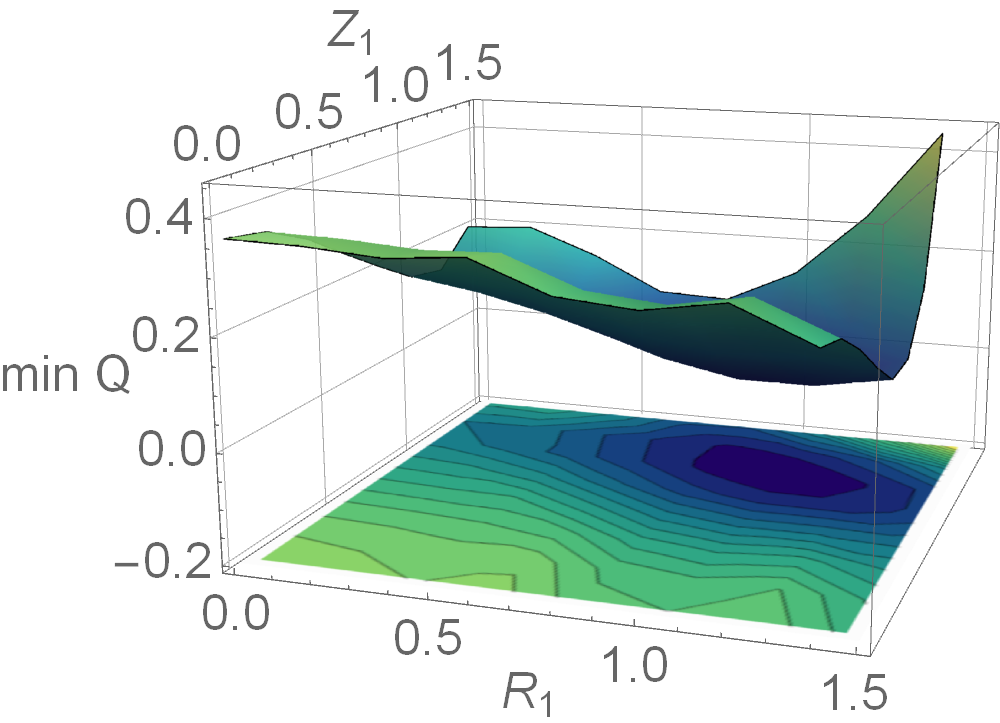}
  \caption{using Differential Evolution - a global optimization routine.}
  \label{fig:New2_global}
\end{subfigure}
\caption{$\min_{R_2,R_3,R_4,R_5} Q(R_1,R_2,R_3,R_4,R_5,Z_1$ with respect to $R_1$ and $Z_1$,}
\label{fig:minimaQ1_2}
\end{figure}

\subsection{Fourier representation of  stellarators}
We now turn to examples of explicit choices of the coefficients $R_{0n},Z_{0n}, \rho_{m,n}$ and $b_n$, corresponding to stellarator plasma boundaries that have been explored in this context in the past. We begin with examples from \cite{Hirshman-1985}, who analysed shapes using spectral condensation, thus providing a convenient point of comparison with this technique. 

\begin{figure}
\centering
\begin{subfigure}{.5\textwidth}
  \centering
  \includegraphics[width=1.\textwidth,trim={4cm 4cm 12cm 4cm},clip]{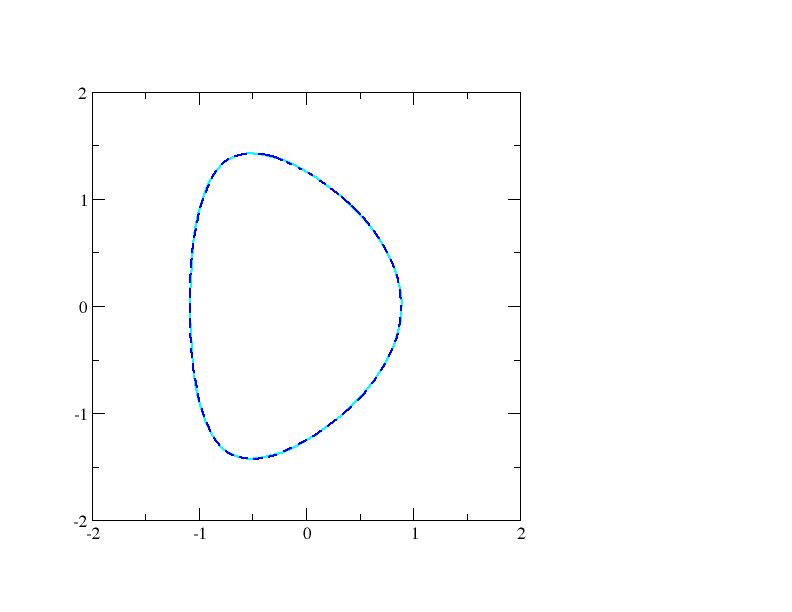}
  \caption{D shape}
  \label{fig:dshape}
\end{subfigure}%
\begin{subfigure}{.5\textwidth}
  \centering
  \includegraphics[width=1.\textwidth,trim={4cm 4cm 12cm 4cm},clip]{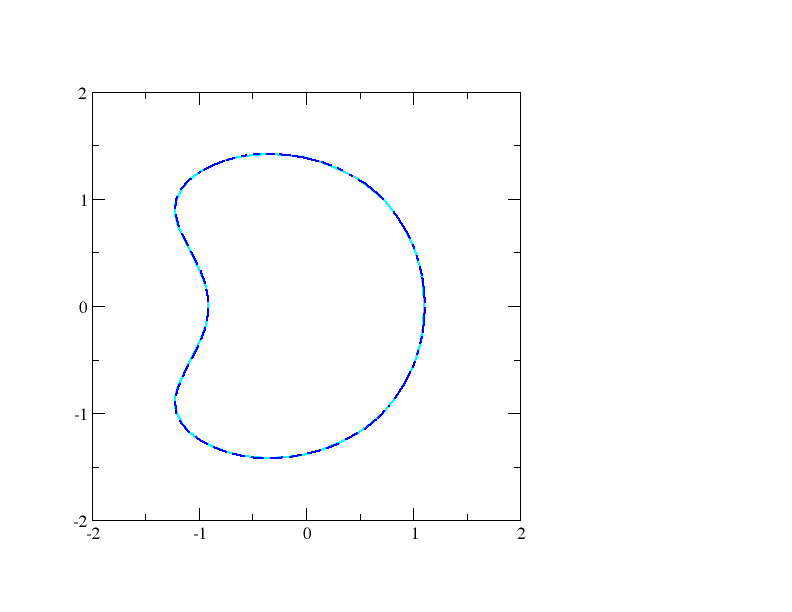}
  \caption{bean shape}
  \label{fig:bean}
\end{subfigure}
\caption{D shape and bean shape reproduced based on~\cite{Hirshman-1985} where the solution of our boundary representation overlaps with the original boundary.}
\label{fig:HirshmanShapes}
\end{figure}

We start with a planar D-shaped boundary given by $R=-0.23+0.989 \cos \theta + 0.137 \cos 2\theta$ and $Z=1.41 \sin\theta - 0.109 \sin 2 \theta$ after spectral condensation \citep{Hirshman-1985}. We use Fourier decomposition to obtain the coefficients in our unique boundary representation that reproduce this boundary, restricting the number of modes $m$ to be such that all the coefficients exceed $0.01$. The result is $R_{00}=-0.306, b_0=1.426$ and $\rho=0.957 \cos \theta + 0.207 \cos 2\theta + 0.032 \cos 3 \theta$. 
The error compared to the original boundary is $\approx 0.4\%$ although the same number of Fourier harmonics are used as in the spectral condensation technique.

Hirshman and Meier also considered a bean-shaped surface given by $R=-0.320+1.115 \cos \theta + 0.383 \cos 2\theta - 0.0912 \cos 3\theta + 0.0358 \cos 4 \theta - 0.0164 \cos 5\theta$ and $Z=1.408 \sin\theta + 0.154 \sin 2 \theta - 0.0264 \sin 3\theta$ after spectral condensation. Applying our representation to this case, we obtain $R_{00}=-0.184, b_0=1.419$ and $\rho=1.184 \cos \theta + 0.208 \cos 2\theta - 0.143 \cos 3 \theta + 0.064 \cos 4\theta - 0.032 \cos 5 \theta + 0.011 \cos 6 \theta$ with an error of $\approx 0.15\%$.
Our representation thus needs even fewer Fourier harmonics than this spectral condensation. 
\footnote{Hirsman and Meier also consider a third case, a so-called belt pinch boundary, which cannot be reproduced by our boundary representation since it has multiple minima and maxima in the vertical coordinate.}

Finally, we consider a representative example from Wendelstein 7-X, where the magnetic field in the so-called standard configuration was calculated using an equilibrium solver in free-boundary mode. As shown in Fig.~\ref{fig:W7-x}, the resulting plasma boundary can be faithfully reproduced with mode numbers $m\leq 5$ and $|n|\leq 3$. 
\begin{figure}%[h!]
    \centering
    \includegraphics[width=.7\textwidth,trim={0 0 7cm 0},clip]{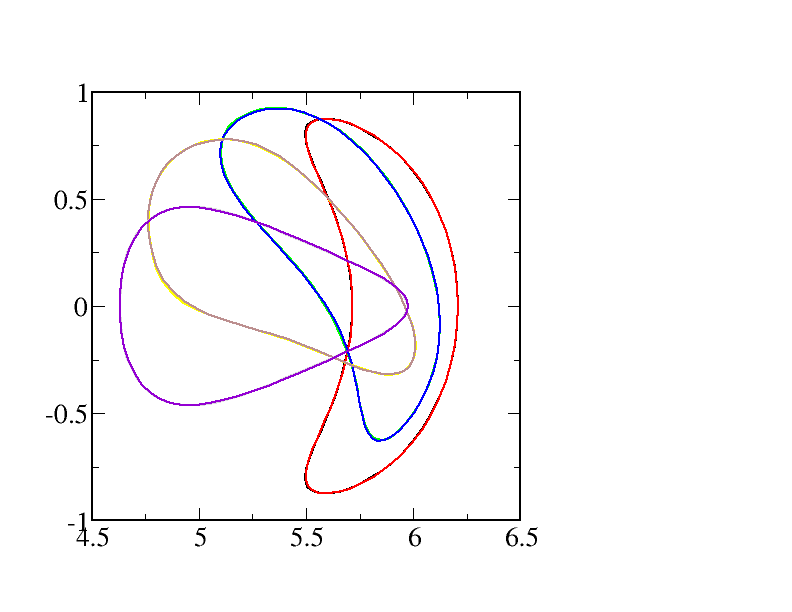}
    \caption{The boundary at different toroidal angle of Wendelstein 7-X. The original overlaps mostly with the replication.}
    \label{fig:W7-x}
\end{figure}

Thus, the representation proposed in Sec.~\ref{Sec:explicitBR} can accurately and economically reproduce relevant plasma boundary shapes, including Wendelstein 7-X and other cases studied earlier in the literature. It does not always need as few Fourier harmonics as spectral condensation, but for ``reasonable'' shapes it appears comparable in efficiency and avoids the need for computational optimization, which is an integral part of the spectral condensation technique. 

\subsection{Application to 3D stellarator optimization}

Finally, we put our boundary representation to the test in a real stellarator optimization problem, where the plasma boundary serves as input for a fixed-boundary equilibrium calculation and is adjusted iteratively until a target function reflecting plasma performance has been minimized. As described in the introduction, such optimization calculations have in the past usually been performed with the ambiguous boundary representation (\ref{conventional representation}). 

We start the optimization with a rotating elliptical boundary, Fig.~\ref{fig:rotatingellipse}.
\begin{figure}%[h!]
    \centering
    \includegraphics[width=.7\textwidth,trim={0 0 7cm 0},clip]{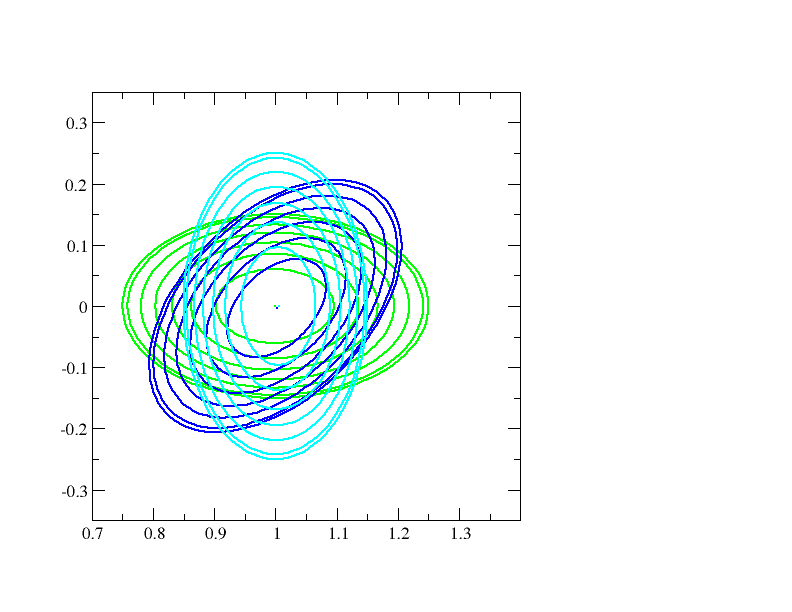}
    \caption{Cross section of rotating ellipse.}
    \label{fig:rotatingellipse}
\end{figure}
and use the optimization code ROSE \citep{ROSE19} with the equilibrium code VMEC and a non-gradient, non-global optimization algorithm (Brent). The target rotational transform is chosen to be 0.25 on axis and 0.35 at the plasma boundary, and in addition we require the magnetic well to exceed a certain threshold (0.1) and the toroidal projection of the plasma boundary to be convex in every point. 

As usual in this type of optimization, the target function $f$ is a weighted sum of squares,
$$
f=\sum_i w_i (F_i-\tilde F_i)^2,
$$
where $F_i$ is the value for criterion $i$, $\tilde F_i$ the corresponding target value, and $w_i$ the $i$'th weight, which can be adjusted to obtain various different optimal (Pareto) points. 

Of course, the performance of the optimization depends of on the exact choice of $w_i$ and $F_i$ as well as the initial condition, but we find that the results turn out much better, and more quickly, with the novel representation than with the standard one used in VMEC. With the same weights chosen for both cases, we typically obtain a penalty value $f$ that is two orders of magnitude smaller when the new boundary representation is employed. The resulting configuration is thus significantly different, and much better, than that obtained with the conventional method. An example of how the plasma boundaries differ is shown in Fig.~\ref{fig:comp}. In this example, all the aims of the optimization were attained when the novel scheme was used, whereas the usual one did not succeed in achieving the prescribed rotational transform and a non-concave plasma boundary. Similar results have also been found with other, more complicated, optimization targets. 

\begin{figure}
\centering
\begin{subfigure}{.5\textwidth}
  \centering
  \includegraphics[width=1.\textwidth,trim={0 0 7cm 0},clip]{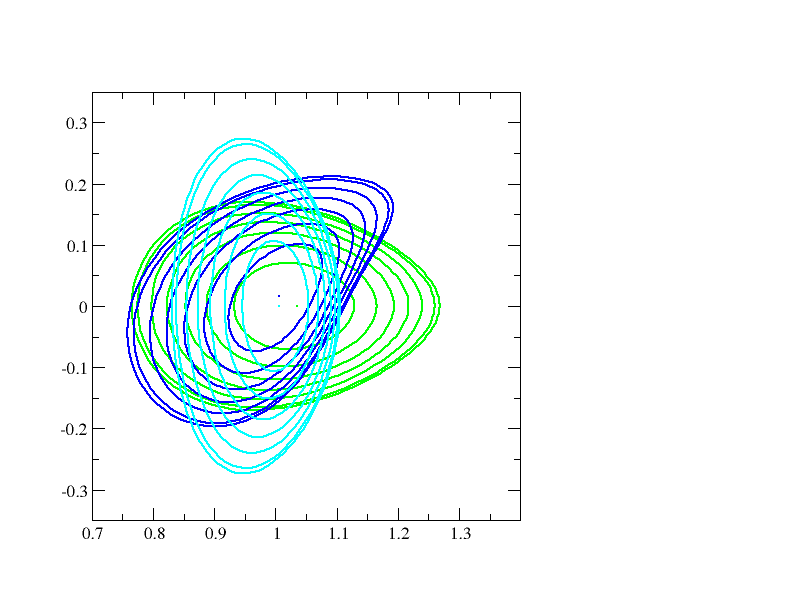}
  \caption{ Cross sections of optimized plasma boundary using standard VMEC boundary representation.}
  \label{fig:non_CPS}
\end{subfigure}%
\begin{subfigure}{.5\textwidth}
  \centering
  \includegraphics[width=1.\textwidth,trim={0 0 7cm 0},clip]{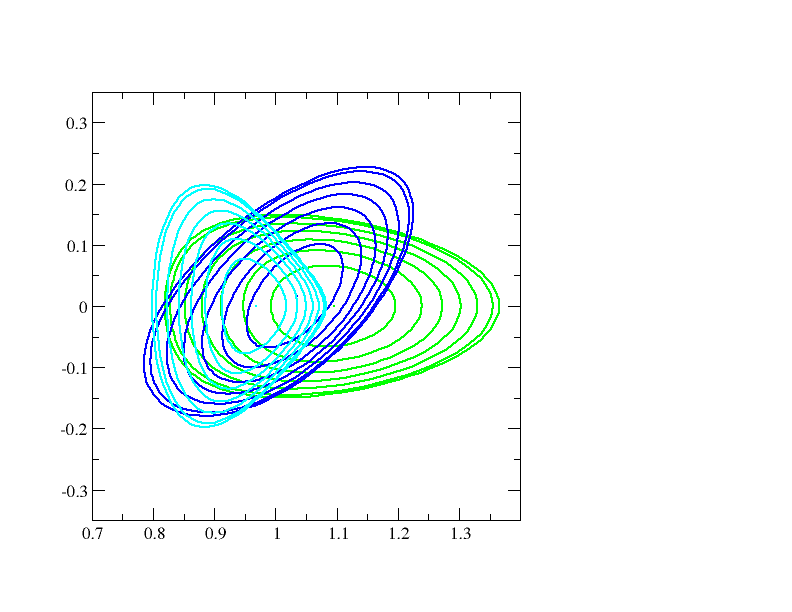}
  \caption{ Cross sections of optimized plasma boundary using unique boundary representation described in Sect.~\ref{Sec:explicitBR}.}
  \label{fig:CPS}
\end{subfigure}
\caption{The poloidal cross sections of optimized plasma boundary and flux surfaces with simple penalty function for the toroidal angles $\varphi=0^{\circ}$ in green, $45^{\circ}$ in dark blue, and $90^{\circ}$ in cyan.}
\label{fig:comp}
\end{figure}

\section{Conclusions}

In summary, the usual Fourier-series representation of the plasma boundary used in stellarator optimization contains much redundancy due to the arbitrariness of the poloidal angle. This redundancy grows exponentially with the number of terms in the series and unnecessarily increases the dimensionality of the search. It causes a plethora of local minima to appear in the optimization landscape, as can be illustrated with simple 2D examples. The situation can be remedied by making the poloidal angle unique, but some care must be taken to ensure that simple stellarator shapes can still be represented in an economical way. When this is done, local minima are eliminated and the optimization proceeds more rapidly than with the usual representation. The outcome also tends to be better, especially if a non-global optimization algorithm is used. 

Our specific boundary parametrization (\ref{Rmn})-(\ref{Zmn}) is simple and intuitive, and requires less computation than the spectral condensation method, but it cannot describe stellarator boundaries with multiple maxima in the $\zeta$-direction. Such boundaries are however highly unusual, and the representation can easily be generalized to include such shapes.

\end{enumerate}

\subsection*{Acknowledgement}
The primary author would like to thank G. Plunk and B. Shanahan for helpful conversations. 
 This work has been carried out in the framework of the EUROfusion Consortium and has received funding from the Euratom research and training programme 2014-2018 and 2019-2020 under grant agreement no. 633053. It was also supported by a grant from the Simons Foundation (560651, PH).The views and opinions expressed herein do not necessarily reflect those of the European Commission.\\

\bibliographystyle{jpp}

\bibliography{per}
\end{document}